\documentclass[prl,aps,nobalancelastpage,amssymb,twocolumn,citeautoscript]{revtex4}

\usepackage{graphicx,multirow}

\begin{document}
\preprint{PRL/Xu \textit{et al}.}

\title{Electronic Nematicity Revealed by Torque Magnetometry in Iron Arsenide EuFe$_2$(As$_{1-x}$P$_x$)$_2$}

\author{ Xiaofeng Xu$^1$, W. H. Jiao$^2$, N. Zhou$^1$, Y. K. Li$^1$, B. Chen$^{3,1}$, C. Cao$^1$, Jianhui Dai$^1$, A. F. Bangura$^4$, Guanghan Cao$^2$}

\affiliation{$^{1}$Department of Physics, Hangzhou Normal University, Hangzhou 310036, China\\
$^{2}$State Key Lab of Silicon Materials and Department of Physics, Zhejiang University, Hangzhou 310027, China\\
$^{3}$Department of Physics, University of Shanghai for Science $\&$ Tehcnology , Shanghai,
China\\
$^{4}$RIKEN(The Institute of Physical and Chemical Research), Wako, Saitama 351-0198, Japan.\\}

\date{\today}

\begin{abstract}
Electronic nematics, an electron orientational order which breaks the underlying rotational
symmetry, have been observed in iron pnictide superconductors several years after their discovery.
However, the universality of the doping dependence of this phase and its relation to other
symmetry-breaking orders (such as superconductivity) in distinct families of iron pnictides, remain
outstanding questions. Here we use torque magnetometry as a probe to study the rotational symmetry
breaking in EuFe$_2$(As$_{1-x}$P$_x$)$_2$ without introducing external pressure. The nematic phase
is found to proliferate well above the structural transition and to persist into the
superconducting regime at optimal doping, after which it becomes absent or very weak, in sharp
contrast to the behaviour observed in BaFe$_2$(As$_{1-x}$P$_x$)$_2$. These measurements suggest a
putative quantum nematic transition near optimal doping under the superconducting dome.
\end{abstract}

\maketitle

The electronic nematic phase, an emergent quantum state in which rotational invariance is broken
spontaneously and the electron fluid exhibits orientational order, has recently attracted a great
deal of attention, largely due to  its close proximity to the superconducting phase in both
high-$T_c$ cuprates and iron
pnictides\cite{Lee06,Daou10,Cooper09,Fradkin10,Chu10,Kasahara12,Chu12,Kuo11,Yi11,Luo13,Fernandes10}.
In cuprates, electronic anisotropy sets in upon opening of the pseudogap, as nicely observed by the
Nernst effect measurements on detwinned YBa$_2$Cu$_3$O$_y$\cite{Daou10}. In iron pnictides, despite
the growing evidence for nematicity borne out by resistivity measurements under stress or
strain\cite{Chu10,Chu12,Kuo11}, neutron scattering\cite{Luo13}, shear
modulus\cite{Fernandes10,Bohmer14}, and Raman spectroscopy\cite{Gallais13}, relatively little is
known about its universality amongst different families or even its microscopic origin.

Torque magnetometry proves to be a powerful thermodynamic tool for studying any anisotropic
susceptibility in materials\cite{Kasahara12,Xu10,Okazaki11}. From a thermodynamic point of view,
magnetic torque is the first derivative of free energy with respect to angular displacement and as
such, tends to be zero in an isotropic material. This is because the torque as given by
$\tau=M\times H$, has $M$ (the magnetization) aligned parallel to $H$ (the applied
field)\cite{footnote1}. The torque therefore only develops whenever any anisotropy sets in. In real
cases, the anisotropy may come from two distinct sources, one from external impurities and the
other from an anisotropic electronic state\cite{Kasahara12}. Considering a situation where the
magnetic field is rotating in the $ab$ plane of a crystal, see the schematic illustration in Fig.
1(a), (the lowest order 2$\phi$ component of) the torque can be written as\cite{Kasahara12}:

\begin{eqnarray}
\tau_{2\phi}&&=Acos2(\phi+\phi_0)\nonumber\\
&&=A_{nem}cos2(\phi+\phi_{nem})+A_{ext}cos2(\phi+\phi_{ext})\label{eqn:one}
\end{eqnarray}

\noindent Here the first and the second terms correspond to the electron nematics and external
contributions, respectively. $A$, $A_{nem}$, $A_{ext}$ denote the amplitudes with initial phase
$\phi_0$, $\phi_{nem}$, $\phi_{ext}$ accordingly. Essentially, the nematic contribution may arise
from two channels, the difference in $\chi_{aa}$ and $\chi_{bb}$, as well as the off-diagonal
$\chi_{ab}$, viz., $A_{nem} cos2(\phi+\phi_{nem} )=\frac{1}{2} H^2 [(\chi_{aa}-\chi_{bb}
)sin2\phi-2\chi_{ab} cos2\phi]$, although it turns out that only $\chi_{ab}$ comes into play in
iron pnictides\cite{Kasahara12}. It is worth noting that $\phi_{nem}$ and $\phi_{ext}$ are both
temperature independent. As a result, the phase $\phi_0$ will solely depend on the relative weight
between $A_{nem}$ and $A_{ext}$. In other words, $\phi_0$ undergoes a significant shift from
$\phi_{ext}$ once the nematic $A_{nem}$ develops a non-zero value. Physically, this is exactly what
was used to identify the nematic phase in BaFe$_2$(As$_{1-x}$P$_x$)$_2$\cite{Kasahara12}.

\begin{figure*}
\includegraphics[width=14cm,keepaspectratio=true]{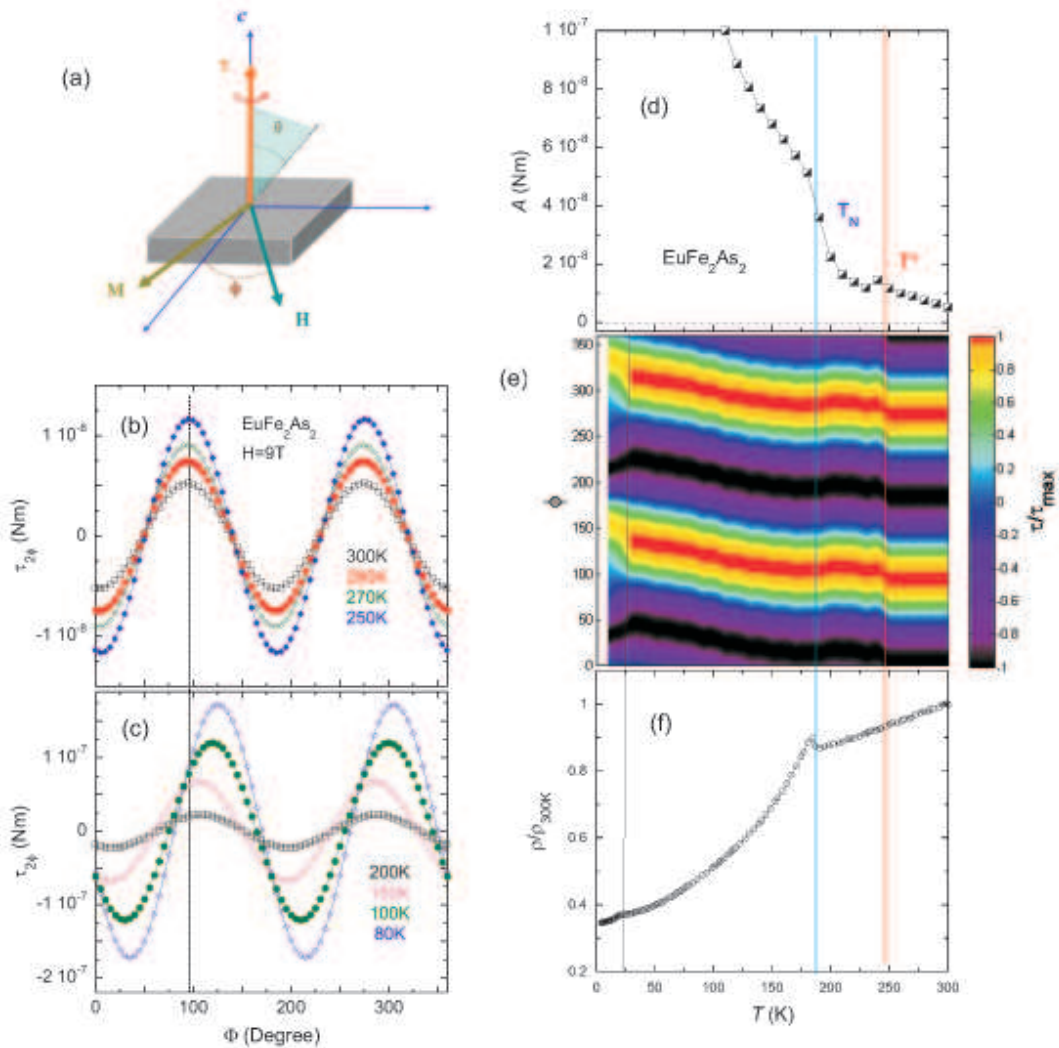}
\caption{(Color online) The angular dependence of magnetic torque with fields rotating in the $ab$
plane of EuFe$_2$As$_2$ parent compound. \textbf{(a)} The schematic diagram for the field
orientation in the setup. Note the $x$- and $y$-axes do not necessarily denote the crystal $a$-axis
and $b$-axis due to the lack of the knowledge about the crystalline axes in the plane which however
will not affect our conclusions in the main text. \textbf{(b)} and \textbf{(c)} show the angular
dependence of the torque above and below the nematic onset temperature $T^*$, respectively.
\textbf{(d)}. The temperature dependence of the amplitude $A$ (see Eqn. (1)). \textbf{(e)}. The
contour plot of the torque as a function of both temperature and angle. Note the torques are
renormalized to the maximal values at each temperature. \textbf{(f)}. The zero-field resistivity
for the strain-free sample.} \label{Fig1}
\end{figure*}

The material in question in this study is a prototypical 122-family iron pnictide
EuFe$_2$(As$_{1-x}$P$_x$)$_2$\cite{Ren09,Jiang09,Jeevan11,Maiwald12,Zapf13}. In the \textit{parent}
EuFe$_2$As$_2$ compound, the high-temperature tetragonal phase undergoes structural and magnetic
transitions coincidentally below $T_s$ (=$T_N$). Upon isovalent P substitution on As sites, both
$T_s$ and $T_N$ are gradually suppressed and superconductivity emerges, reaching a maximum
superconducting transition temperature $T_c$ of $\sim$30 K at optimal doping level of
$x$$\sim$0.2\cite{Jeevan11,Maiwald12}. This gives a phase diagram akin to the one generic to 122
family. In addition, the rare earth Eu$^{2+}$ develops a magnetic ordering at low temperature
$\sim$20 K . Intriguingly, this magnetic order evolves with P doping level, antiferromagnetically
(AFM) coupled at low P concentrations and ferromagnetically (FM) on the overdoped
side\cite{Jeevan11,Zapf13}.

Here we use the temperature evolution of the angular magnetic torque to detect the possible nematic
phase above $T_s$ in EuFe$_2$(As$_{1-x}$P$_x$)$_2$ (see Supplementary Information (SI) for
experimental details). The initial phase of the magnetic torque $\phi_0$ was seen to be
significantly shifted below $T^*$, a new temperature scale signifying the onset of the nematic
order, which is well above $T_s$ and $T_N$. This nematic order was found to decrease with
increasing P concentrations up to the optimal doping and to become very weak (or even absent) in
the overdoped region. The resultant phase diagram suggests a possible quantum nematic transition
beneath the superconducting dome, which in turn may have an intimate connection with the observed
superconductivity.

\begin{figure*}
\includegraphics[width=14cm,keepaspectratio=true]{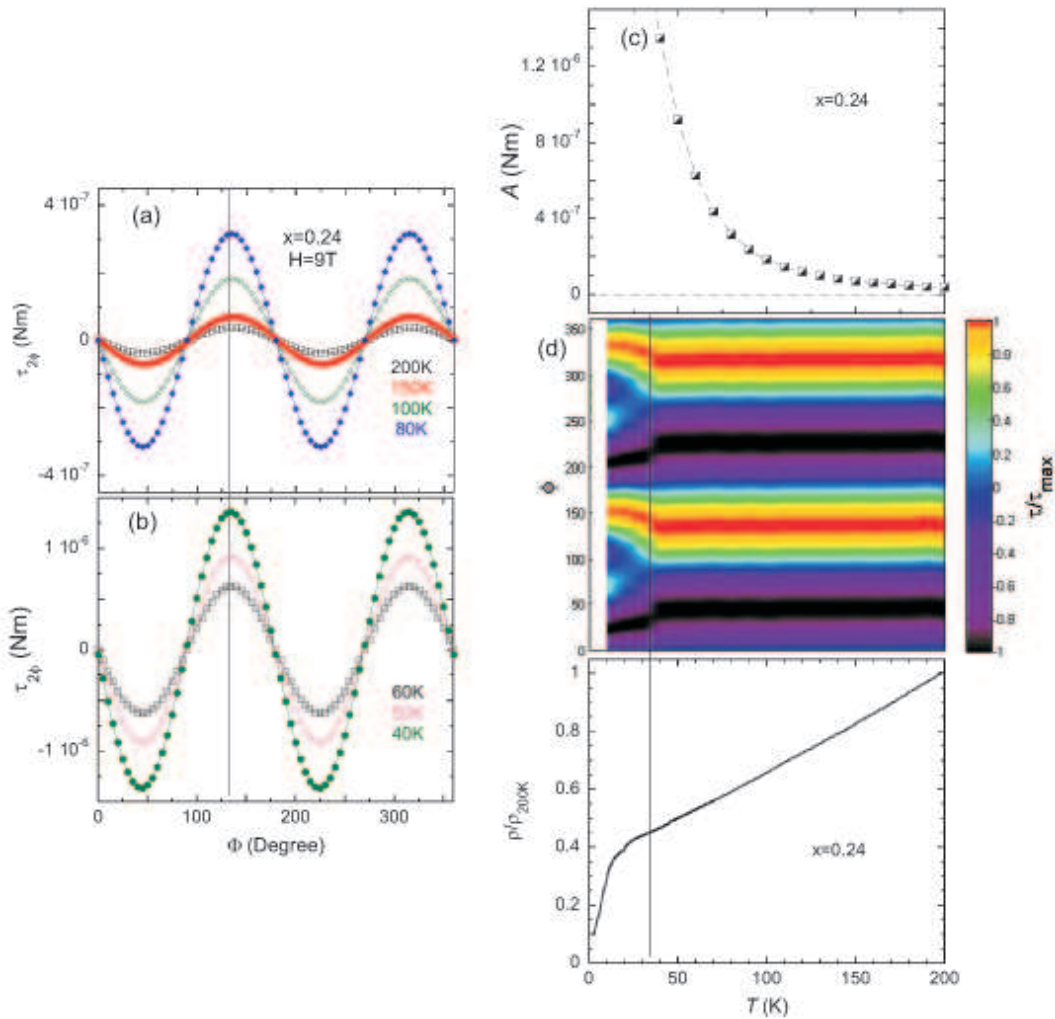}
\caption{(Color online) The angular dependence of the torque for fields rotating in the $ab$ plane
of the overdoped sample ($x$=0.24). \textbf{(a)} and \textbf{(b)}. The angular torque at different
temperatures. \textbf{(c)} summarizes the resultant amplitude $A$ at each temperature studied.
\textbf{(d)}. The contour plot of the renormalized torque as a function of temperature and angle.
\textbf{(e)} presents the zero-field resistivity for strain-free sample.} \label{Fig2}
\end{figure*}

The panels (b) and (c) in Figure 1 show the representative angular torque at various fixed
temperatures under 9 tesla field for the parent compound EuFe$_2$As$_2$. Clearly, a magnetic torque
is manifest even at 300 K, indicative of the considerable anisotropy of the magnetization at room
temperature. With decreasing temperatures, the sinusoidal torque gets larger while kept at a
constant phase. However, this angular profile alters significantly below $\sim$250 K, as shown in
Fig. 1(c). The phase is no longer \textit{locked} into the high-$T$ one and evidently seen to vary
with temperatures. This fundamental change of the angular torque is best visualized in a contour
plot presented in Fig. 1(e) where the data are renormalized to their maximal values at each
individual temperature, $\tau/\tau_{max} (T)$. The temperature dependence of the amplitude $A$ (see
Eqn. (1)) and the zero-field resistivity $\rho(T)$ for a stress-free sample were also incorporated
as Fig. 1 (d) and (f), respectively. In spite of the smooth $\rho(T)$ curve across $\sim$250 K, a
new temperature scale, marked as $T^*$, manifests itself as a kink in $A(T)$ in Fig. 1(d) and the
considerable change of the angular torque profile in Fig. 1(e). The structural transition at
$T_s$($\sim$$T_N$) $\sim$180 K, displays in Fig. 1 (d) as a sharp increase in $A(T)$, in Fig. 1(e)
as a inflexion point, and in Fig. 1 (f) as an upturn in the resistivity. As the temperature is
lowered down to $\sim$20 K, $\rho(T)$ undergoes a moderate kink, associated with the magnetic order
of Eu$^{2+}$ ions. This magnetic order also has profound effects on the torque data (Fig. 1 (e)) by
introducing a large 4-fold angular component (see SI).

The above phenomenon can overall be understood in the preceding dichotomy analysis. At high
temperatures, only the torque from external impurities takes a role therefore it will maintain the
same phase $\phi_{ext}$. As the sample is cooled below $T^*$, a second nematic contribution sets in
such that the phase will shift to $\phi_0$  which is distinct from the high-$T$ $\phi_{ext}$.
Apparently, this new phase $\phi_0$ depends on the relative weights of the two contributions. In
this sense, $T^*$ corresponds to the onset temperature for the electronic nematic phase.

In order to confirm that only the $ab$ plane is responsible for the above findings, we perform the
field rotation out of the FeAs plane (data shown in SI). Interestingly, the phase of this
angle-dependent torque barely changes with temperature, in stark contrast to the case of in-plane
field rotation. Collectively, these provide compelling evidence for the nematic origin of the phase
shift observed upon in-plane rotation.

Similarly, we also studied the in-plane angular evolution of the magnetic torque in underdoped
($x$=0.18, $T_c$=21 K) and nearly optimally-doped ($x$=0.2, $T_c$=29 K) samples. Similar phase
shifts were also observed in these two doping levels, although at much lower temperatures. The
onset temperatures for the nematic phase, $T^*$, were identified as $\sim$140 K and $\sim$100 K,
respectively (see SI). Although $T^*$ for the optimally-doped sample remains high, the relative
phase shift below $T^*$ gets much smaller, indicative of the weak nematicity.

We next consider an overdoped sample ($x$=0.24, determined from EDX, see SI) whose resistivity,
Fig. 2(e), starts an incipient dip at $\sim$20 K and a quick drop below $\sim$14 K but non-zero
resistivity is observed down to 2.5 K, indicating either sample inhomogenity or strong internal
field induced by Eu$^{2+}$ FM ordering on the overdoped side\cite{Ren09,Jeevan11}. Angular torque
in the normal state (Fig. 2 (a) and (b)) is sinusoidal and maintains the same phase all the way
down to the superconducting (fluctuation) temperature, below which a substantial 4-fold component
develops (see SI for the torque in the superconducting state). It is noted that this 4-fold
symmetry torque sets in about 10$\sim$20 K above $T_c$, similar to what was observed in $x$=0.18
and $x$=0.2 samples. We attribute this to the superconducting fluctuations which have the same
effects on the torque as the Eu$^{2+}$ order in the parent compound. Overall, this locking of the
torque phase with temperature is vividly captured in the contour plot, Fig. 2 (d). On the other
hand, the temperature dependence of the amplitude, $A(T)$ in Fig. 2 (c), evolves in a smooth manner
in the normal state.

Figure 3 shows the revised phase diagram of EuFe$_2$(As$_{1-x}$P$_x$)$_2$ revealed from our torque
measurements. In addition to the phases uncovered thus far by other measurements, we have found a
rather broad region above the structural and magnetic transitions where the electrons in the FeAs
plane start to develop a novel, orientational order that breaks the rotational invariance of the
underlying tetragonal lattice and reduce the symmetry from \textit{C}$_4$ to
\textit{C}$_2$\cite{Fradkin10}. This nematic phase is found to extend all the way up to the optimal
doping where the structural and magnetic transitions are believed to be completely suppressed. On
the overdoped side, however, no nematic phase can be revealed above the superconducting transition,
in contrast to the phase diagram of BaFe$_2$(As$_{1-x}$P$_x$)$_2$ where nematicity clearly survives
in the very overdoped region\cite{Kasahara12}. This indicates that the phase diagram associated
with the nematic order is not universal, even within the 122-family.

\begin{figure}
\includegraphics[width=9cm,keepaspectratio=true]{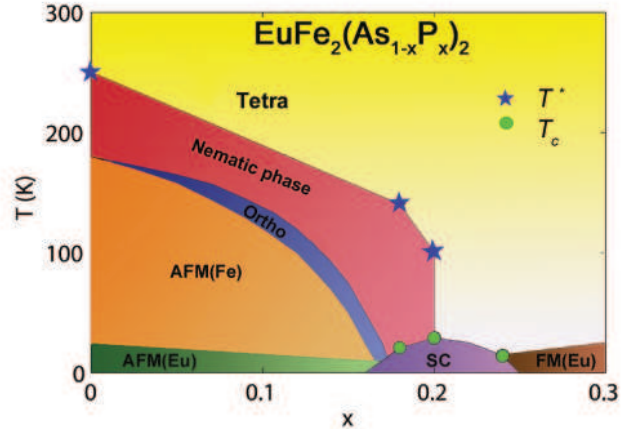}
\caption{(Color online) The resultant phase diagram of EuFe$_2$(As$_{1-x}$P$_x$)$_2$ as derived
collectively from our measurements and the previous studies. For simplicity, we neglect the fine
spin structure of Eu$^{2+}$, including that in the superconducting state, which were recently
uncovered in Ref. \cite{Zapf13,Nandi14}.} \label{Fig3}
\end{figure}

We note that the electronic anisotropy above the structural and magnetic transitions has also been
investigated by a thermoelectric power (TEP) study in three specimens of
EuFe$_2$(As$_{1-x}$P$_x$)$_2$, two non-superconducting samples ($x$=0.05 and 0.09) and one
overdoped sample ($x$=0.23)\cite{Jiang13}. Similarly, the nematic phase had only been detected in
the $x$=0.05 and $x$=0.9 samples, no anisotropy being observed in the overdoped $x$=0.23 sample.
Remarkably, for $x$=0.05, the anisotropy in the TEP appears to develop even above $\sim$250 K, a
temperature we assigned as $T^*$ for the onset temperature of the nematicity in the parent compound
(Note that the TEP was performed under a uniaxial stress clamp. It may effectively enhance the
nematicity\cite{Jiang13}). Consistently, on the overdoped side, no nematic order can be detected in
the TEP measurements nor our magnetic torque study. For the TEP measurements, it is difficult to
define the onset temperature $T^*$ since the uniaxial pressure is necessary to detwin the sample.
However, thanks to the unbalanced twin-domain volumes, torque measurements prove to be an effective
approach to study any anisotropy in a stress-free sample\cite{Kasahara12}.

It is unlikely that the absence of the nematicity on the overdoped side is due to the sample
inhomogeneity. First, the non-zero resistivity below $T_c$ in our sample does not necessarily imply
the sample inhomogeneity as the internal field induced by Eu$^{2+}$ FM order in overdoped sample
may be comparable to the upper critical field. Second, no nematicity has either been detected by
the TEP study in the overdoped sample, whose sample homogeneity is not a serious issue
there\cite{Jiang13}. Moreover, it is noteworthy that even under the uniaxial stress, no resistivity
anisotropy has been observed in the overdoped 122 samples, including Co and Ni doped
BaFe$_2$As$_2$\cite{Kuo11}.

A natural question raised by our study concerns the relation between these various transitions in
the phase diagram and the microscopic origin of the nematicity. As argued in Ref.
\cite{Kasahara12}, because \textit{C}$_4$ rotational symmetry can only be broken once, the
structural distortion in the diagram is not a real phase transition, but rather a transition dubbed
'meta-nematic transition' in which the order parameter has a sharp increase \textit{albeit} it is
non-zero on either side of the transition. Instead, the nematic transition is the genuine
second-order transition because the order parameter (proportional to $A_{nem}$ in the torque) is
only non-zero below $T^*$.  This is further confirmed by the divergent nematic susceptibility, and
the nematic order is considered as the driving force behind the structural instability\cite{Chu12}.

The origin of this electronic anisotropy is highly controversial to date and two alternative
proposals have been widely discussed. In the Ising-nematic scenario\cite{Fang08,Xu08,Fernandes11,
Fernandes12,Fernandes13}, the nematic phase is characterized by a broken \textit{Z}$_2$ Ising
symmetry (spins differentiate in the otherwise equivalent directions and are apt to point along one
of the in-plane axes) and is driven by spin fluctuations above the AFM order of Fe. In the other
theory\cite{Lee09,Lv09,Chen10}, it is the ferro-orbital ordering, uneven population of the $d_{xz}$
and $d_{yz}$ orbitals, that gives rise to the nematicity which ultimately makes the orthorhombic
crystal lattice energetically more stable. For the present system, the nematic order clearly
survives to the optimal doping where the long-range AFM order is thought to be completely
suppressed. However, quantum spin fluctuations can still in principle induce the nematicity in this
regime. How the quantum fluctuations play a role in the highly overdoped
BaFe$_2$(As$_{1-x}$P$_x$)$_2$, warrants furthur theoretical insights. Alternatively, if the orbital
ordering is the driving force\cite{Fernandes13}, the termination of the nematic phase near optimal
doping is somehow puzzling. Whether the disappearance of the nematic phase is related to a Lifshitz
transition near this doping region remains to be seen\cite{Jiang13,Thirupathaiah11}. It is worth
noting that recent TEP experiments seemingly disagree with the orbital-ordering proposal as the
origin of nematicity because it does not match the sign of the TEP anisotropy above
$T_s$\cite{Jiang13}. The contrasting doping dependence of the nematic phase in 122-families also
imposes stringent challenges on any theory which relies on orbital-ordering.

To conclude, magnetic torque measured in EuFe$_2$(As$_{1-x}$P$_x$)$_2$ reveals a new phase boundary
associated with nematic phase formation up to the optimal superconducting transition. This nematic
transition is not present in the overdoped side of the phase diagram, heralding a possible quantum
(nematic) transition near optimal doping concentration\cite{Fernandes13a}. Indeed, quantum critical
points have been uncovered in iron pnictides, in particular for the 122-families, near to the
optimal doping\cite{Jiang09JPCM,Dai09,Hashimoto12,Walmsley13,Analytis14}. Future studies are
therefore needed to clarify whether the nematic fluctuations associated with this quantum nematic
transition are the fundamental ingredient for its enhanced
superconductivity\cite{Fernandes13b,Fernandes13c}.

We thank N. E. Hussey, C. M. J. Andrew, C. Lester, Xiaofeng Jin for stimulating discussions. This
work is sponsored by the NSFC (Grant No. 11104051, 11104053).

\bibliography{SrPt2As2}

\begin{thebibliography}{21}

\bibitem{Lee06} P. A. Lee, N. Nagaosa, X. G. Wen,  Rev. Mod. Phys. {\bf 78}, 17 (2006).
\bibitem{Daou10} R. Daou \textit{et al.}, Nature. {\bf 463}, 519 (2010).
\bibitem{Cooper09} R. A. Cooper \textit{et al.}, Science {\bf 323}, 603 (2009).
\bibitem{Fradkin10} E. Fradkin, S. A. Kivelson, M. J. Lawler, J. P. Eisenstein, and A. P. Mackenzie, Annu. Rev. Condens. Matter Phys. {\bf 1}, 153 (2010).
\bibitem{Chu10} J. H. Chu, J. G. Analytis, K. De Greve, P. L.  McMahon, Z. Islam, Y. Yamanoto, and I. R. Fisher, Science {\bf 329}, 824 (2010).
\bibitem{Kasahara12} S. Kasahara \textit{et al.}, Nature {\bf 486}, 382 (2012).
\bibitem{Chu12} J. H. Chu, H. H. Kuo, J. G. Analytis, and I. R. Fisher, Science {\bf 337}, 710 (2012).
\bibitem{Kuo11} H. H. Kuo \textit{et al.}, Phys. Rev. B {\bf 84}, 054540 (2011).
\bibitem{Yi11} M. Yi \textit{et al.}, Proc. Natl Acad. Sci. USA {\bf 108}, 6878 (2011).
\bibitem{Luo13} H. Q. Luo \textit{et al.}, Phys. Rev. Lett. {\bf 111}, 107006 (2013).
\bibitem{Fernandes10} R. M. Fernandes \textit{et al.}, Phys. Rev. Lett. {\bf 105}, 157003 (2010).
\bibitem{Bohmer14} A. E. B\"{o}hmer \textit{et al.}, Phys. Rev. Lett. {\bf 112}, 047001 (2014).
\bibitem{Gallais13} Y. Gallais \textit{et al.}, Phys. Rev. Lett. {\bf 111}, 267001 (2013).
\bibitem{Xu10} X. Xu \textit{et al.}, Phys. Rev. B {\bf 81}, 224435 (2010).
\bibitem{Okazaki11} R. Okazaki \textit{et al.}, Science {\bf 331}, 439 (2011).
\bibitem{footnote1} The torque has a zero value when fields point along high-symmetry directions (e.g., the $c$-axis) where the free energy has extreme value, recalling the torque is the first derivative of the free energy to angle.
\bibitem{Ren09} Z. Ren \textit{et al.}, Phys. Rev. Lett. {\bf 102}, 137002 (2009).
\bibitem{Jiang09} S. Jiang \textit{et al.}, New J. Phys. {\bf 11}, 025007 (2009).
\bibitem{Jeevan11} H. S. Jeevan, D. Kasinathan, H. Rosner, and P. Gegenwart,  Phys. Rev. B {\bf 83}, 054511 (2011).
\bibitem{Maiwald12} J. Maiwald, H. S. Jeevan, and P. Gegenwart,  Phys. Rev. B {\bf 85}, 024511 (2012).
\bibitem{Zapf13} S. Zapf \textit{et al.}, Phys. Rev. Lett. {\bf 110}, 237002 (2013).
\bibitem{Nandi14} S. Nandi \textit{et al.},  Phys. Rev. B {\bf 89}, 014512 (2014).
\bibitem{Jiang13} S. Jiang, H. S. Jeevan, J. Dong, and P. Gegenwart, Phys. Rev. Lett. {\bf 110}, 067001 (2013).
\bibitem{Fang08} C. Fang, H. Yao, W. F. Tsai, J. Hu, and S. A. Kivelson,  Phys. Rev. B {\bf 77}, 224509 (2008).
\bibitem{Xu08} C. Xu, M. M\"{u}ller, and S. Sachdev, Phys. Rev. B {\bf 78}, 020501(R) (2008).
\bibitem{Fernandes11} R. M. Fernandes, E. Abrahams, and J. Schmalian, Phys. Rev. Lett. {\bf 107}, 217002 (2011).
\bibitem{Fernandes12} R. M. Fernandes, A. V. Chubukov, J. Knolle, I. Eremin, and J. Schmalian, Phys. Rev. B {\bf 85}, 024534 (2012).
\bibitem{Fernandes13} R. M. Fernandes, A. V. Chubukov, J. Schmalian, Nat. Phys. \textbf{10}, 97 (2014).
\bibitem{Lee09} C. C. Lee, W. G. Yin, and W. Ku, Phys. Rev. Lett. {\bf 103}, 267001 (2009).
\bibitem{Lv09} W. Lv, J. Wu, and P. Phillips, Phys. Rev. B {\bf 80}, 224506 (2009).
\bibitem{Chen10} C. C. Chen \textit{et al.}, Phys. Rev. B {\bf 82}, 100504 (2010).
\bibitem{Thirupathaiah11} S. Thirupathaiah \textit{et al.}, Phys. Rev. B {\bf 84}, 014531 (2011).
\bibitem{Fernandes13a} R. M. Fernandes, S. Maiti, P. W\"{o}lfle and A. V. Chubukov, Phys. Rev. Lett. {\bf 111}, 057001 (2013).
\bibitem{Jiang09JPCM} S. Jiang \textit{et al.}, J. Phys. Condens. Matter {\bf 21}, 382203 (2009).
\bibitem{Dai09} J. Dai \textit{et al.}, Proc. Natl Acad. Sci. USA {\bf 106}, 4118 (2009).
\bibitem{Hashimoto12} K. Hashimoto \textit{et al.}, Science {\bf 336}, 1554 (2012).
\bibitem{Walmsley13} P. Walmsley \textit{et al.}, Phys. Rev. Lett. {\bf 110}, 257002 (2013).
\bibitem{Analytis14} J. G. Analytis \textit{et al.}, Nat. Phys. doi:10.1038/nphys2869.
\bibitem{Fernandes13b} R. M. Fernandes, and A. J. Millis, Phys. Rev. Lett. {\bf 111}, 127001 (2013).
\bibitem{Fernandes13c} R. M. Fernandes, and A. J. Millis, Phys. Rev. Lett. {\bf 110}, 117004 (2013).





\end{thebibliography}


\end{document}